\documentclass{aa}
\usepackage{graphics}
\usepackage{epsf}

\newcommand{\oversim}[2]{\protect{\mbox{\lower0.5ex\vbox{%
   \baselineskip=0pt\lineskip=0.2ex
   \ialign{$\mathsurround=0pt #1\hfil##\hfil$\crcr#2\crcr\sim\crcr}}}}} 
\newcommand{\simless} {\mbox{$\,\mathrel{\mathpalette\oversim<}\,$}} % <~ sign
\newcommand{\sig}{\:\lower0.6ex\hbox{$\stackrel{\textstyle >}{\sim}$}\:}
\newcommand{\sil}{\:\lower0.6ex\hbox{$\stackrel{\textstyle <}{\sim}$}\:}
\newcommand{\sigs}{\:\lower0.4ex\hbox{$\stackrel{\scriptstyle
      >}{\scriptstyle \sim}$}\,}
\newcommand{\sils}{\:\lower0.4ex\hbox{$\stackrel{\scriptstyle
      <}{\scriptstyle \sim}$}\,}

\begin{document}

\title{The mean surface density of companions in a
    stellar-dynamical context}
    \author{Ralf Klessen{\inst{1,2,3}}
    \and Pavel Kroupa{\inst{4}}
}
\institute{UCO/Lick Observatory, University of California,  499 Kerr Hall, 
    Santa Cruz, CA 95064, USA (e-mail: ralf@ucolick.org)  \and Otto Hahn Fellow,
    Max-Planck-Institut f{\"u}r Astronomie, K{\"o}nigstuhl 17, 69917
    Heidelberg, Germany \and Sterrewacht Leiden, Postbus 9513, 2300-RA Leiden,
    The Netherlands \and
    Institut f\"{u}r Theoretische Physik und Astrophysik, Universit\"at
    Kiel, 24098 Kiel, Germany (e-mail: pavel@astrophysik.uni-kiel.de)}
\offprints{R.\ Klessen}
\date{}
\thesaurus{05.03.1; 08.02.3; 08.06.2; 10.15.1}
\titlerunning{The mean surface density of companions}
\authorrunning{Ralf Klessen and Pavel Kroupa}
\maketitle

\begin{abstract}
Applying the mean surface density of companions, $\Sigma(r)$, to the
dynamical evolution of star clusters is an interesting approach to
quantifying structural changes in a cluster.  It has the advantage
that the entire density structure, ranging from the closest binary
separations, over the core-halo structure through to the density
distribution in moving groups that originate from clusters, can be
analysed coherently as one function of the stellar \mbox{separations $r$}.

This contribution assesses the evolution of $\Sigma(r)$ for clusters
with different initial densities and binary populations. The changes
in the binary, cluster and halo branches as the clusters evolve are
documented using direct $N$-body calculations, and are correlated with
the cluster core and half-mass radius. The location of breaks in the
slope of $\Sigma(r)$ and the possible occurrence of a binary gap can
be used to infer dynamical cluster properties.

\keywords{binaries: general -- open clusters and
associations: general -- stars: formation -- stellar dynamics}
\end{abstract}

\section{Introduction}
\label{sec:introduction}
Studying the clustering properties of stars in star-forming regions is
a necessary input towards understanding the formation and evolution of
young stellar clusters (Elmegreen \& Efremov 1997). Using two-point
angular correlation functions to analyse the spatial distribution of
stars, Gomez et al.\ (1993) showed that young stars in the
Taurus-Auriga molecular cloud form in small associations containing of
the order of ten stellar systems.  Larson (1995) extended this
investigation by taking into account the results from different
searches for binary companions to the pre-main sequence stars in the
Taurus-Auriga region.  He computed the mean surface density of
companions, $\Sigma(\theta)$, per star as a function of angular
separation $\theta$.  This statistical measure is closely related to
the two-point correlation function but does not require
normalisation. As two different power laws are necessary to fit the
data, with a slope $\approx -2$ for separations below $\approx
0.04\,$pc and with a slope $\approx -0.6$ above, Larson concluded that
there are two distinct clustering regimes. At small separations, the
derived companion density is determined by binaries and higher-order
multiple systems, whereas at large separations the overall spatial
structure of the stellar cluster is observed. Larson interpreted the
observed non-integer slope in the clustering regime as evidence for
fractal structure of the cluster.  In addition, Larson noted that the
break of the distribution occurs in Taurus-Auriga at length scales
which are equivalent to the typical Jeans length in molecular clouds,
corresponding to a Jeans mass of $\approx 1\,$M$_{\odot}$. He
speculated that stellar systems with smaller separations form from the
fragmentation of single collapsing proto-stellar cores, whereas the
spatial distribution on larger scales is due to the hierarchical
structure of the parent molecular cloud.

These results prompted the reanalysis of the Taurus-Auriga data by
other authors (Simon 1997, Bate, Clarke \& McCaughrean 1998, Gladwin
et al.\ 1999), as well as the subsequent investigation of additional
star-forming regions.  Mean surface densities of companions as a
function of angular separation have been derived for Orion (Simon
1997, Bate et al.\ 1998, Nakajima et al.\ 1998), for the
$\rho$-Ophiuchus cloud (Simon 1997, Bate et al.\ 1998, Nakajima et
al.\ 1998, Gladwin et al.\ 1999), and for the star-forming regions in
Chamaeleon, Vela and Lupus (Nakajima et al.\ 1998).  Common to all
studies is that the companion surface density is best described as a
double power-law, with slopes of $\approx -2$ in the binary branch and
slopes between $-0.9$ and $-0.1$ in the large-scale clustering
regime. However, the length scales where the break of the distribution
is found to vary considerably, from $\approx 400\,$AU in the Trapezium
cluster in Orion, over $\approx 5\,\!000\,$AU in Ophiuchus and
$\approx 8\!\,000\,$AU in Taurus, to $30\,\!000\,$AU for the Orion OB
region.  This fact raises considerable doubts about the interpretation
of the break location as being determined by the Jeans conditions in
the cloud. This would imply quite different Jeans masses which in turn
should lead to deviations of the initial mass function, which have not
been observed.

A thorough theoretical evaluation of the mean surface density of
companions, $\Sigma(\theta)$, and a discussion of viable
interpretations can be found in Bate et al.\ (1998).  Altogether the
following picture emerges: At small separations, $\Sigma(\theta)$
traces the separations of binary stars and higher-order multiple
stellar systems. The slope $\approx -2$ results from the frequency
distribution of binary separations being roughly uniform in logarithm
(Duquennoy \& Mayor 1991, for main sequence stars). The break occurs
at the `crowding' limit, i.e.\ at separations where wide binaries
blend into the `background' density of the cluster. At larger
separations, $\Sigma(\theta)$ simply reflects the large-scale spatial
structure of the stellar cluster. Bate et al.\ (1998) pointed out that
$\Sigma(\theta)$ can be strongly affected by boundary effects and that
a non-integer power-law slope in the cluster branch does not
necessarily imply fractal structure. They showed that a simple
core-halo structure, as is typical for evolved stellar clusters, will
result in a non-integer slope for separations larger than the core
radius.  They also speculated about possible effects of dynamical
cluster evolution on the properties of $\Sigma(\theta)$. 

It is the aim of the present paper to investigate, for the first time,
evolutionary effects on $\Sigma(\theta)$ as derived from realistic
$N$-body computations. We use models studied by Kroupa (1995a,b,c,
1998, hereinafter K1-K4) for a comparison with a `standard' dynamical
analysis, where the binary population is analysed separately from the
bulk cluster properties. For comparison, the mean surface density of
companions for models of {\em proto}stellar clusters that form and
evolve through turbulent molecular cloud fragmentation is discussed
in Klessen \& Burkert (2000, 2001).

The structure of the paper is as follows. In the next section
(\S\ref{sec:msdc}) we mathematically define the mean surface density
of companions, $\Sigma(r)$, and briefly discuss its limitations. In
\S\ref{sec:models} we describe the star cluster models and their
properties. In \S\ref{sec:evolutionary-effects}, we investigate the
influence of cluster evolution on $\Sigma(r)$ and in particular
discuss wide-binary depletion. The effects of averaging and projection
are analysed in \S\ref{sec:statistical-effects}, and a possible
observational bias is discussed in \S\ref{sec:observational-bias}.
\S\ref{sec:morphology} discusses features in $\Sigma(\theta)$ and
their relation to cluster morphology.  Finally, our results are
summarised in \S\ref{sec:summary}

\begin{table*}[t]
  \begin{center}
    \begin{tabular}[h]{rlcccccc}
\hline
&&Model ${\cal A}$ &Model ${\cal B}$ &Model ${\cal C}$ &Model ${\cal D}$ &Model ${\cal E}$ &Model ${\cal F}$ \\
\hline
$n$ & & 5 & 5 & 5  & 5 & 3 & 3 \\
 $f$ & & 1 & 1 & 1 & 1 & 0 & 0 \\
 $r_{\rm h}$ & (pc) & 0.08 & 0.25 & 0.8 & 2.5 & 0.08 & 0.25 \\
 $\tau_{\rm h}$ & ($10^6\,$yr) & 0.094 & 0.54 & 3.0 &
17 & 0.094 & 0.54 \\
 log$_{10}\,\rho_0$ & (stars/pc$^{3}$) & 5.6 &4.1 &2.7 &1.1 &5.6 &4.1 \\
 log$_{10}\,\Sigma_0$ & (stars/pc$^{2}$) & 4.3 & 3.3 & 2.3 & 1.2 & 4.3 & 3.3 \\
\hline
$t=t_{\rm init}$ & ($10^6\,$yr)  & 0.0  & 0.0  & 0.0  & 0.0 & 0.0  & 0.0  \\
$t=12.5\,\tau_{\rm h}$ & ($10^6\,$yr)   & 1.19 & 6.67 & 37.5 & 211 & 1.19 & 6.68 \\
$t=125\,\tau_{\rm h}$ & ($10^6\,$yr)   & 11.9 & 66.7 & ---  & --- & 11.9 & 66.8 \\
$t=t_{\rm end}$ & ($10^6\,$yr)    & 297  & 300  & 300  & 296 & 297  & 300  \\
\hline
    \end{tabular}
    \end{center}
    \hfill
    \vspace{0.2cm}
    \parbox[b]{\textwidth}{
    \caption[tab1]{ Properties of the model clusters. All clusters have
    $N=400$ stars with a mean stellar mass of $0.32\,$M$_\odot$, 
    $n$ lists the number of different realizations of the model, and
    $f$ gives the starting binary fraction. Initially, the
    density profile of all clusters follows a Plummer law with
    half-mass radius $r_{\rm h}$, leading to a half-mass diameter
    crossing time $\tau_{\rm h}$, and to central and central surface
    densities $\rho_0$ and $\Sigma_0$, respectively. The times, 
    when we analyse the
    clusters, are indicated in the last four lines.  }
    \label{tab:models} 
    }
\end{table*}
\section{Mean Surface Density of Companions}
\label{sec:msdc}
The mean surface density of companions, $\Sigma(\theta)$, specifies
the average number of neighbours per square degree on the sky at an
angular separation $\theta$ for each cluster star. Knowing the
distance of the cluster, the angular separation $\theta$ between two
stars in the cluster translates into an absolute distance $r$. In the
numerical models we have full access to all phase space coordinates
and we define, in what follows, the mean surface density of companions
$\Sigma(r)$ as the number of stars per pc$^2$ as a function of the
projected distance $r$ (in pc). In Section~\ref{subsec:projection} we
show that the results are invariant to which plane is used for
projection.  We calculate for each star $i$ in the system the
projected distance $r_{ij}$ to all other stars $j \ne i$. The
separations $r_{ij}$ are sorted into annuli with radii $r$ and width
$\delta r$, where we use logarithmic binning such that each decade in
separation is partitioned into 10 logarithmically equidistant bins
($\log_{10} \delta r = 0.1$).  To obtain the function $\Sigma(r)$, we
divide the number $\delta\!\!\;N(r)$ of stellar pairs per annulus $r$
by the surface area $2\pi r\delta r$ and average by dividing by the
total number $N$ of stars in the cluster. The mean surface density of
companions as a function of separation $r$ then follows as
\begin{equation}
\Sigma(r) \equiv \frac{\delta\!\!\;N(r)}{2\pi N r\delta r}\,.
\end{equation}
The function $\Sigma(r)$ is related to the two-point correlation
function, $\xi(r)$, by $\xi(r) = \left(\Sigma(r) /\langle \Sigma
\rangle \right)-1$, where $\langle \Sigma \rangle$ is the mean stellar
surface density in the considered area (Peebles 1993). Because the
normalisation $\langle \Sigma \rangle$ is often difficult to
determine, it is preferable to use the function $\Sigma(r)$.

Stellar surveys have finite area and boundary effects may occur. For
stars closer than a distance $r_b$ to the boundary all annuli with
$r>r_b$ extend beyond the limits of the survey and companion stars may
be missed. This has little effect when considering small separations
$r$ as only a small fraction of all stars is affected, however, when
$r$ becomes close to the survey size the missing companions result in
a steep decline of $\Sigma(r)$. Several methods have been proposed to
correct for that effect (Bate et al.\ 1998, and references therein),
where either the accepted range of $r$ is reduced to separations much
smaller than the survey size, or additional assumptions about the
background density of stars are made. Neither approach is completely
satisfactory. 

In the present study we do not attempt to adopt any of the correction
methods for the following reasons. First, in the numerical simulations
all information about the system is accessible. There are no
observational constraints and we always consider {\em all} stars in
the cluster. The survey area can be arbitrary large and is chosen such
that it includes the complete cluster. Second, although the considered
clusters are subject to the tidal field of the Galaxy we do not
include Galactic-field stars in consideration. Hence, there is no
confusion limit, where contamination with foreground or background
stars becomes important.

\section{Star Cluster Models}
\label{sec:models}

A range of star-cluster models are constructed, and their dynamical
evolution is calculated using {\sc Nbody5} (Aarseth 1999), which
includes a standard Galactic tidal field (Terlevich 1987).  The
cluster models have been discussed extensively in K1--K4, so that only
a short outline is provided here.

The stellar systems initially follow a Plummer density distribution
(Aarseth, H\'enon \& Wielen 1974) with half-mass radius $r_{\rm h}$, and
the average stellar mass is independent of the radial distance, $r$,
from the cluster centre. 

Stellar masses are distributed according to the solar-neighbourhood
IMF (Kroupa, Tout \& Gilmore 1993) with $0.1\le m\le
1.1\,$M$_\odot$. Larger masses are omitted so as to avoid
complications arising from stellar evolution.  Binaries are created by
pairing the stars randomly, giving a birth binary proportion $f=N_{\rm
bin}/(N_{\rm sing}+N_{\rm bin})$, where $N_{\rm sing}$ and $N_{\rm
bin}$ are the number of single-star and binary systems,
respectively. The initial mean system mass is $2\langle m \rangle$,
with $\langle m \rangle =0.32\,$M$_\odot$ being the average stellar
mass.  This results in an approximately flat mass-ratio distribution
at birth (figure~12 in K2).  Periods and eccentricities are
distributed following K1. The initial periods range from $10^3$ to
$10^{7.5}$~days, and the eccentricity distribution is thermal, i.e.\
the relative number of binaries increases linearly with eccentricity
being consistent with observational constraints.

The parameters are listed in Table~\ref{tab:models}. Four clusters
with $f\!=\!1$ are constructed spanning a wide range of central densities,
from log$_{10}\rho_0=1.1$ to~5.6~[stars/pc$^3$].  Each model contains
$N=400$ stars and has a mass $M_{\rm cl}=128\,$M$_\odot$.  The initial
tidal radius in all cases is $r_{\rm t}\approx8$~pc. All stars are
kept in the calculation to facilitate binary-star analysis, but those
with $r\gg r_{\rm t}$ experience unphysical accelerations in the
linearised local tidal field and rotating coordinate system (Terlevich
1987), so that the density distribution of stellar systems at large
radii does not reflect the true distribution in the moving group.
Five different renditions are calculated for each model to increase
the statistical significance of our results. In addition, two clusters
with $f\!=\!0$ are constructed for comparison with the binary-rich cases.
The evolution of these models is calculated for three cluster
realisations each. The computations cover $1\times10^9\,$years, but we
consider only a sub-set of all possible snap-shots in the current
analysis.

\begin{figure*}[ht]
\unitlength1cm
\begin{picture}(16,15)
\put( 2.0, 0.6){\resizebox{15cm}{!}{\includegraphics{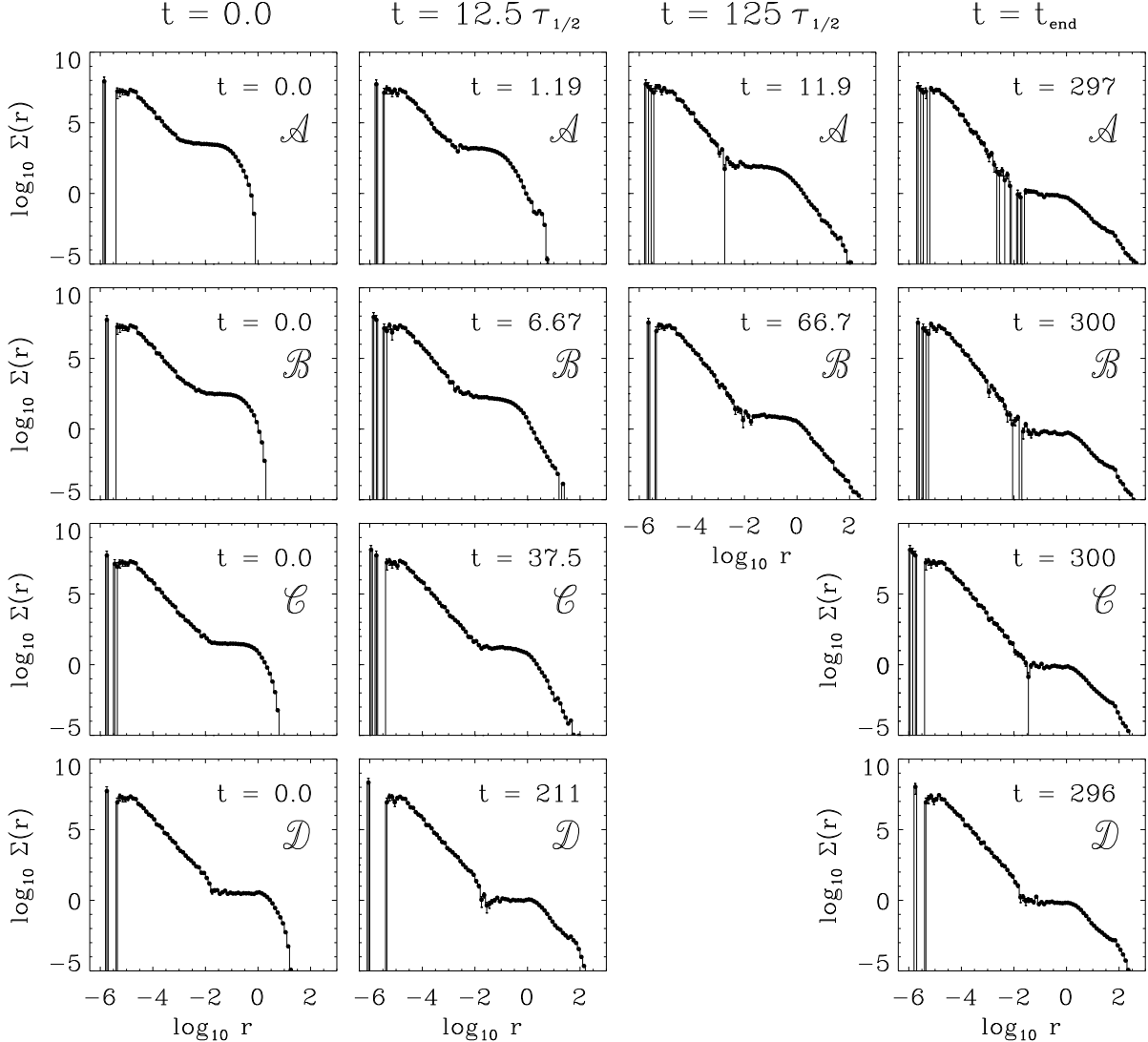}}}
\end{picture}
\hfill
\parbox[t]{\textwidth}{
\caption[fig01]{{\label{fig:AtoD} The mean surface
density of companions, $\Sigma(r)$, as a function of separation $r$ for
star clusters $\cal A$ to $\cal D$ with an initial binary fraction of
100$\,$\% at different states of the dynamical evolution: initially
($t=0.0$, left column), at $t=12.5\,\tau_{\rm h}$ and
$t=125\,\tau_{\rm h}$ (2.\ and 3.\ column) and at $t=t_{\rm end}$ (right
column). The corresponding time in units of $10^6\,$years is indicated
in the upper right corner of each plot.  $\Sigma(r)$ is obtained as an
average over $n=5$ different cluster realizations for each model as a
projection into the $xy$-plane. The error bars indicate Poisson
errors. }}}
\end{figure*}

\section{Evolutionary Effects}
\label{sec:evolutionary-effects}

In this section, we discuss the influence of the dynamical cluster
evolution on the resulting mean surface density of companions
$\Sigma(r)$. For each star cluster, $\Sigma(r)$ is calculated as an
average over the set of $n$ individual model realisations, and we restrict
ourselves to discussing the projection into the $xy$-plane. The
influence of averaging and projection is discussed in
\S\ref{sec:statistical-effects}.

\subsection{The Global Evolution of Stellar Clusters}
\label{subsec:density}
The evolutionary sequence of $\Sigma(r)$ for models $\cal A$ to $\cal
D$ is illustrated in figure \ref{fig:AtoD}. Its first column denotes
the initial state of each system at $t = 0$, the second and third
columns depict the function $\Sigma(r)$ taken at $t= 12.5\,\tau_{\rm
h}$ and $t= 125\,\tau_{\rm h}$, where $\tau_{\rm h}$ is the initial
half-mass diameter crossing time (table~1).  As $\tau_{\rm h}$
increases with increasing $r_{\rm h}$, these equivalent stages of
cluster evolution correspond to different absolute times as indicated
in the figure. The final state of the systems, at roughly $t\approx
3\times10^8\,$yr, is given in the last column. For models $\cal C$ and
$\cal D$, $125\,\tau_{\rm h} > t_{\rm end}$, so that the third column
in figure \ref{fig:AtoD} contains no entry.
\begin{figure*}[th]
\unitlength1cm
\begin{picture}(16,8.5)
\put( 2.0, 0.6){\resizebox{15cm}{!}{\includegraphics{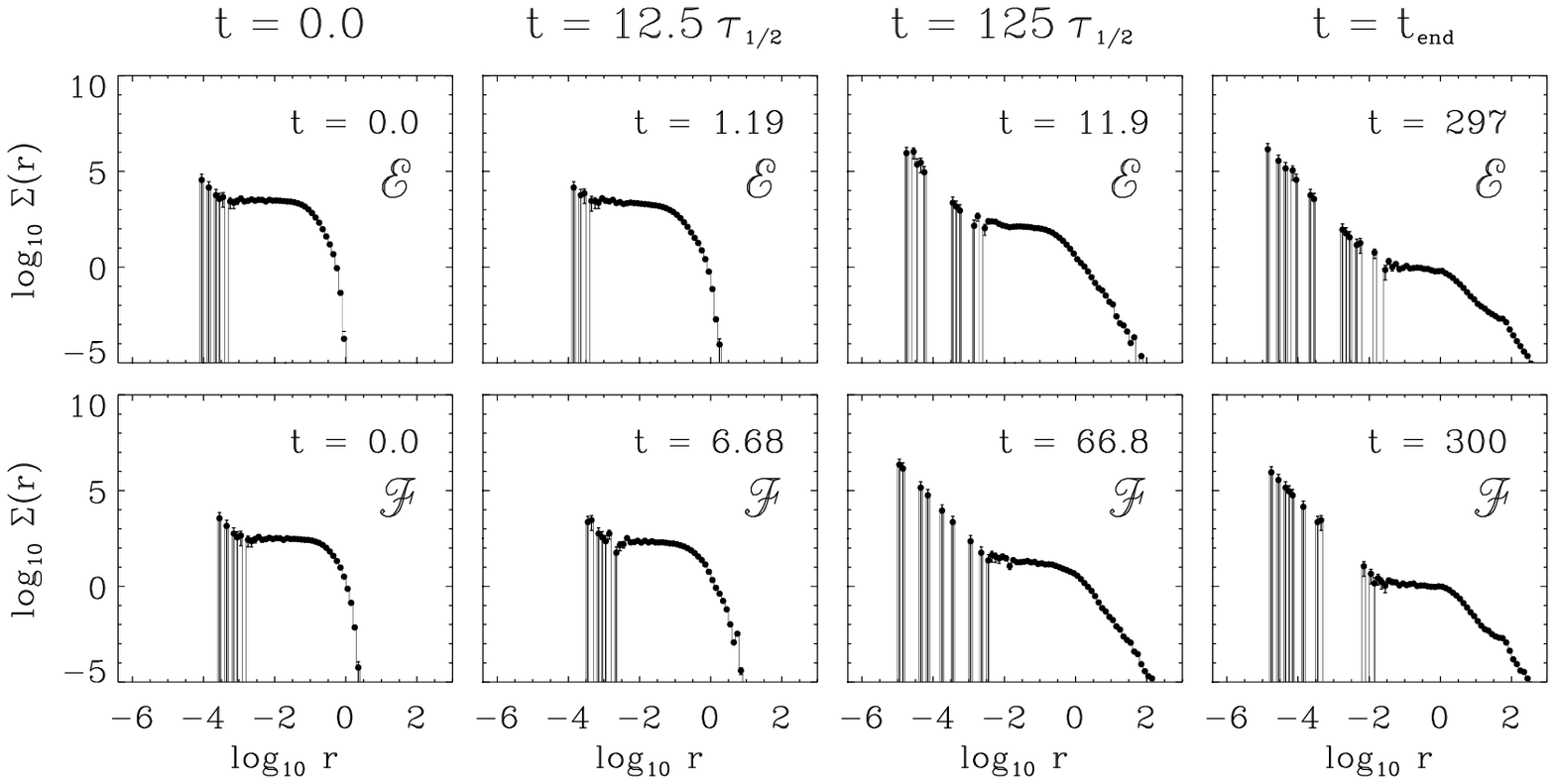}}}
\end{picture}
\hfill
\parbox[t]{\textwidth}{
\caption[fig02]{{\label{fig:EandF} The mean surface
density of companions, $\Sigma(r)$, as a function of separation $r$ for
star clusters $\cal E$ and $\cal F$. These models are equivalent to
$\cal A$ and $\cal B$, respectively, but initially contain no binary
stars. The figure is analogous to figure \ref{fig:AtoD}, except that
$n=3$ different realizations of each model are used in the averaging
process. }}}
\end{figure*}

At any time and for all models, the mean surface density of companions
exhibits, at low separations $r$, a well defined power-law behaviour
$\Sigma(r) \propto 1/r^2$. It implies an approximate uniform
distribution of binary separations in $\log r$ (Bate et al.\ 1998).
For larger $r$, the binary branch blends into the plateau of constant
companion density corresponding to the core of the star cluster. This
{\em first} break of the distribution occurs at separations $r_{1}$
where the number of chance projections of cluster members becomes
equal to the number of binaries in that separation bin. At larger $r$
chance projections completely dominate $\Sigma(r)$, and wide binaries,
if in fact present, can no longer be identified as such. The location
of the first break therefore depends on the binary fraction and on the
central density of the cluster.  On larger scales the stellar clusters
follow a Plummer radial density profile with half-mass radius $r_{\rm
h}$. Therefore, a {\em second} break occurs at $r_{2}$ and $\Sigma(r)$
declines sharply for separations $r \sig r_{\rm h}\approx r_{2}$ (see
Section~\ref{sec:morphology}).

As the dynamical evolution progresses, the clusters expand and the
density declines. Hence, the projected mean surface density of
companions decreases as well. While many of the binaries with
separations comparable to the mean distance of stellar systems in the
cluster core become disrupted, some new binaries may form by capture.
Usually these are higher-order multiples (K2) with separations close
to the first break or smaller. As the cluster expands, the binary
branch becomes less affected by crowding and the first break in
$\Sigma(r)$ shifts to greater separations.  This behaviour is clearly
visible in figure \ref{fig:AtoD}. For all models the core plateau in
$\Sigma(r)$ `decreases in height' and `moves' to larger separations as
time progresses. At late stages of the evolution the entire binary
branch is uncovered and a few long-period orbits appear through
capture. This is also documented in figs.~3 and~4 in K4.

The clusters develop core-halo structures through energy
equipartition. Low-mass stars gain kinetic energy through encounters
with more massive stars. The low-mass stars move away from the cluster
centre, forming the halo, whereas the more massive stars sink towards
the centre. The trajectories of halo stars that trespass beyond
the tidal radius of the cluster are dominated by the Galactic tidal
field, and most become unbound.  Hence, the clusters expand until they fill
their tidal radii.  When this stage is reached (roughly after
$t>1\times10^8\,$years), the different clusters evolve identically
(see also figure 1 in K3), mostly loosing stars through their first and
second Lagrange points (Terlevich 1987; Portegies Zwart et al. 2001).
As a result, $\Sigma(r)$ extends beyond the second break to
increasingly larger separations with an increasingly shallower slope,
which we identify as the halo branch in $\Sigma(r)$.  Near the tidal
radius $r_{\rm t}$, a third break occurs, as stars with $r>r_{\rm
  t}$ become unbound. The trajectories of stars belonging to these
unbound moving groups are not followed with sufficient resolution (see
Section~\ref{sec:models}). Also, these stars are likely to be
severely contaminated by field stars in the Galaxy, and we refrain
from a further discussion of this outermost branch in $\Sigma(r)$.

As can be seen from figure~\ref{fig:AtoD}, knowledge of the initial
global properties of the system is effectively erased through the
dynamical evolution.  The cluster and halo branches in $\Sigma(r)$
look quite indistinguishable in the final frames.  The situation
changes, however, when considering the binary branch, as discussed in
Section~\ref{subsec:binaries}.  The $f\!=\!1$ versus $f\!=\!0$
experiments demonstrate that there is no significant difference in
bulk cluster evolution between clusters containing a large primordial
binary proportion and no binaries (K3).  This is also evident by
studying $\Sigma(r)$; concentrating only on the cluster and halo
branches, the upper two final panels in figures \ref{fig:AtoD} and
\ref{fig:EandF} are indistinguishable.

\begin{figure*}[ht]
\unitlength1cm
\begin{picture}(16,15)
\put( 2.0, 0.6){\resizebox{15cm}{!}{\includegraphics{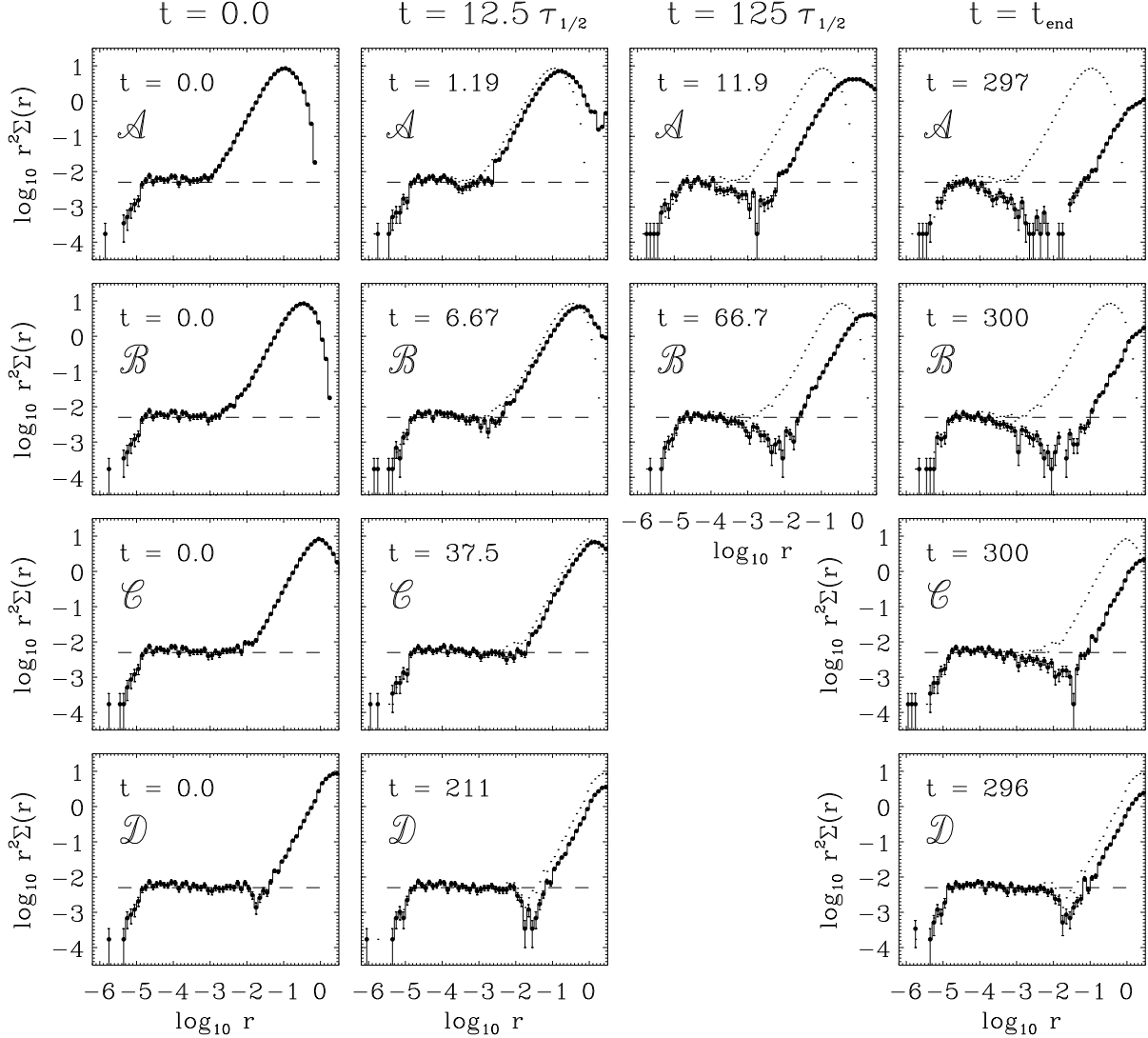}}}
\end{picture}
\hfill
\parbox[t]{\textwidth}{
\caption[fig03]{{\label{fig:gap} The function
$r^2\,\Sigma(r)$ for models $\cal A$ to $\cal D$. The depicted times,
averaging and projection are analogous to figure \ref{fig:AtoD}. The
plots concentrate on the properties of the binary branch, where the
horizontal dashed lines indicate its initial slope. The depletion of
the distribution at late stages in the interval
$10^{-4}\,$pc$\,<\,r\,<\,0.1\,$pc is the result of wide-binary
disruption.  The rising part of $r^2\,\Sigma(r)$ at separations
greater than the first break corresponds to the cluster core, and the
following decline for $r>r_{\rm h}$ is the contribution from the cluster
halo.  To demonstrate the effects of dynamical evolution, the dotted
lines for $t>0.0$ indicate the initial distribution.}}}
\end{figure*}
\subsection{Binary Stars}
\label{subsec:binaries}
Binary systems and wide hierarchical systems form through
capture during the evolution of the clusters, as seen in figure
\ref{fig:EandF}. These systems result from triple or higher-order
stellar encounters, and are consequently very rare.  The periods of
these systems range from $10^7$ to $10^{11}\,$days (figure 10 in K2),
and a well distinguished binary branch with slope $\approx -2$
develops, extending to radii well beyond the initial position of the
first break.

However, as the clusters evolve, binaries are not only created but
also destroyed. Due to their smaller binding energies, wide binaries
are more vulnerable to dynamical processes than close ones. If the
initial binary fraction is high, then the destruction processes
dominate over binary formation, and cluster evolution leads to a
depletion of wide binaries.  As a result, $\Sigma(r)$ steepens on the
large-separation side of the binary branch. In extreme cases, some
annuli $r$ of $\Sigma(r)$ may become completely depopulated, and
consequently a gap between the binary and the cluster branches opens
up, as is noticeable in figure \ref{fig:AtoD}.

The efficiency of binary disruption depends strongly on the initial
density of the cluster. For high stellar densities, the typical
impact parameters of stellar encounters are small. Hence, there is a
relatively high frequency of encounters for which the energy exchange
exceeds the binding energy of typical binary systems, which
subsequently dissolve. In our suite of models, the effect of binary
depletion is largest in $\cal A$, which has the highest central density
$\rho_0$, and decreases with increasing half-mass radius $r_{\rm h}$ as
$\rho_0$ becomes smaller. 

This is demonstrated in figure \ref{fig:gap}. Unlike the previous
figures it shows a reduced range of separations, concentrating on the
binary branch, and it plots $r^2\,\Sigma(r)$ to make it easier to
determine the power-law slope and deviations from it.  At $t=0.0$, for
all models $r^2\,\Sigma(r)$ is constant in the binary branch. This
reflects the initial distribution of the binary separations, which is
uniform in the logarithm within the range $10^{-5}\,$pc to
$10^{-2}\,$pc. Values $r<10^{-5}\,$pc come from the projection of the
3-dimensional distribution into the $xy$-plane.  For model $\cal D$
the complete binary branch is visible and fully segregated from the
cluster branch (which corresponds to the rising part of the
plot). This is because the half-mass radius of the cluster is large
enough that the projected mean separation between cluster members
exceeds the separation of the widest binary system.  As both branches
are clearly separated initially, dynamical evolution does not alter
the binary distribution significantly. There is little sign of wide
binary depletion, even at $t=t_{\rm end}$.

\begin{figure*}[t]
\unitlength1cm
\begin{picture}(16,11.5)
\put( 1.5, -6.4){\resizebox{20cm}{!}{\includegraphics{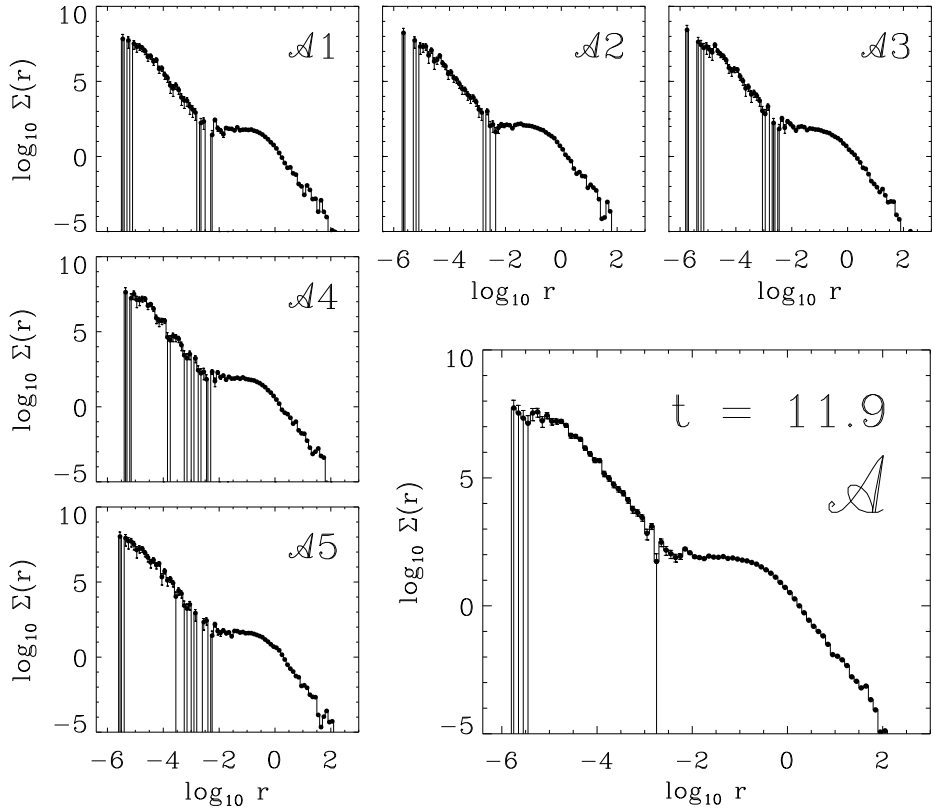}}}
\end{picture}
\hfill
\parbox[t]{\textwidth}{
\caption[fig04]{\footnotesize{\label{fig:variance} Mean surface
density of companions, $\Sigma(r)$, of cluster $\cal A$ at time
$t=125\,\tau_{\rm h}$ for the projection into the $xy$-plane. The figure
illustrates the effect of the averaging process. The small plots show
$\Sigma(r)$ for the five different realizations ${\cal A}1$ to ${\cal A}5$
of model $\cal A$, whereas the large plot gives the resulting averaged
function.  }}}
\end{figure*}

In the other models, the typical separations in the cluster core are
smaller than $10^{-2}\,$pc, and binary and cluster branches overlap in
the beginning. This is most significant for model $\cal A$, where the
initial central density is highest. Consequently a large number of
wide binaries are disrupted during the dynamical evolution of the
system, and $r^2\,\Sigma(r)$ drops considerably below its initial
value (indicated by the dashed line) in the range
$10^{-4}\,$pc$\,<\,r\,<\,10^{-1}\,$pc. Because the size of this gap
depends on the age and the initial central concentration of the
cluster, analysing the signatures in $\Sigma(r)$ could be used to
constrain the initial state of observed stellar clusters. This fact,
namely that the binary population retains a memory of its past
dynamical environment, is also used in K1 to infer the typical
structures in which most Galactic-field stars form, by studying the
shape of the binary period distribution (``inverse dynamical
population synthesis'').

\section{Statistical Effects}
\label{sec:statistical-effects}
In the previous section we analysed the mean surface density of
companions averaged over several model realisations seen in one
projection (into the $xy$-plane). Both, the effect of averaging and
the effect of projection shall be discussed here.
\begin{figure*}[t]
\unitlength1cm
\begin{picture}(16, 9.5)
\put(-1.5, -7.0){\resizebox{22cm}{!}{\includegraphics{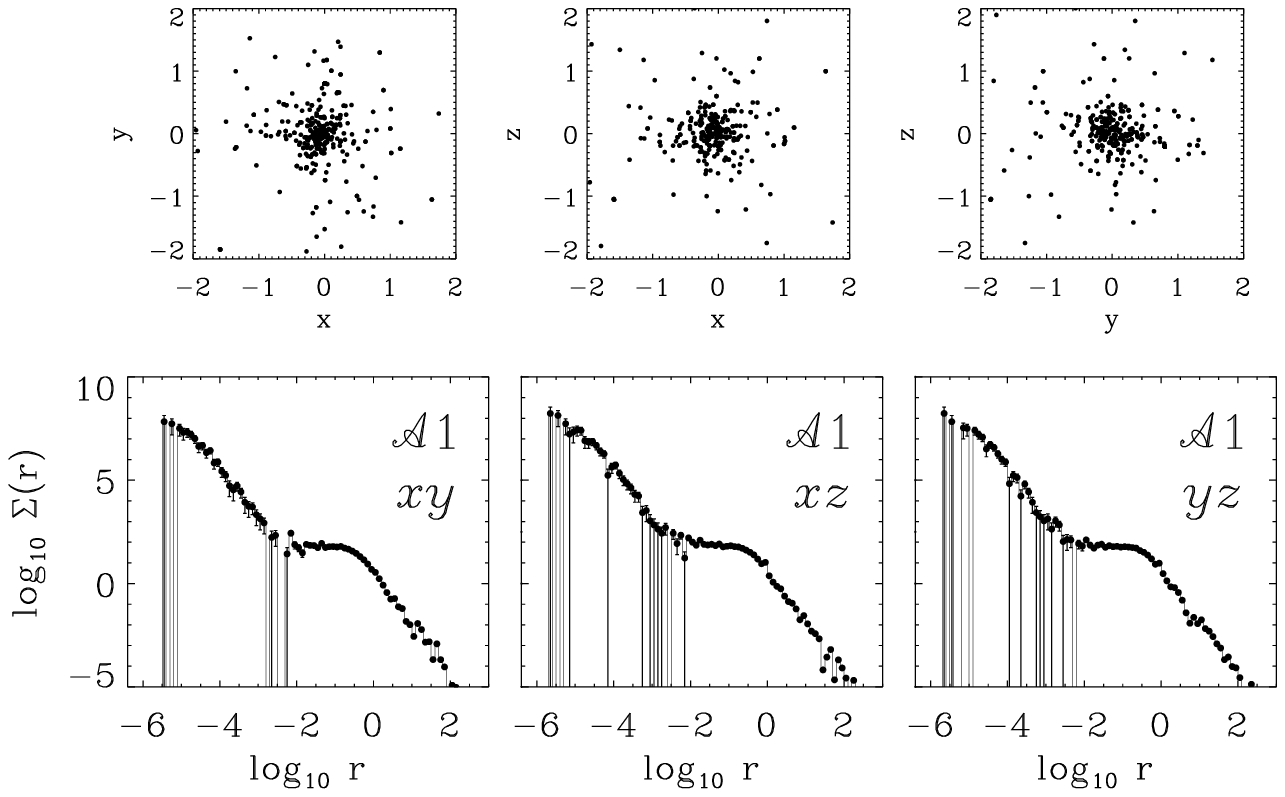}}}
\end{picture}
\hfill
\parbox[t]{\textwidth}{
\caption[fig05]{{\label{fig:projection} Illustration of
the effect of projection.  Mean surface density of companions
$\Sigma(r)$ for cluster realization ${\cal A}1$ at time
$t=125\,\tau_{\rm h}$ for three different projections.  The upper panel
shows the stellar distribution within $\pm 2\,$pc of the cluster
centre and the lower panels show the resulting
$\Sigma(r)$. The global features of $\Sigma(r)$ are independent of
projection.  }}}
\end{figure*}

\subsection{Cluster to Cluster variations}
\label{subsec:averaging}
To estimate the effect of statistical variations between different
model realisations, figure \ref{fig:variance} plots $\Sigma_i(r)$
constructed for each of the $n=5$ individual cluster rendition of
model ${\cal A}$ at $t=125\,\tau_{\rm h}$.  This model is chosen,
because it has the smallest initial crossing time, hence, our
simulations span the largest evolutionary interval.  After the system
has evolved for $125\,\tau_{\rm h}$, differences become noticeable at
separations where the number of neighbours is small.  This is the case
at very large separations, where the stellar density rapidly
decreases, and deviations may also occur at the extreme ends of the
binary branch. At the smallest separations, $\Sigma(r)$ is determined
by only one or two very close binary systems, and near the first break
of the distribution, the depletion of wide binary systems becomes
noticeable and a gap may open up.  As can be seen in figure\
\ref{fig:variance}, the exact location and number of depleted
separation bins slightly varies between different model realizations.
Therefore, the wide binary gap appears wider and more noticeable for
individual realisations compared to the ensemble average.

\subsection{Different Projections}
\label{subsec:projection}
Clusters of young stars are seen on the sky in only {\em one}
projection.  Inferring the full 3-dimensional structure of the cluster
is therefore in principle impossible without additional information or
assumptions. In the previous sections, we concentrated on the
projection into the $xy$-plane when plotting $\Sigma(r)$. For the
spherical clusters considered in the current analysis the different
projections are equivalent, and each gives a fair representation of
the complete system. Only at distances comparable to the tidal radius
does the flattening through the Galactic tidal field of the cluster
become significant (Terlevich 1987).

The invariance to changes in projection is demonstrated in figure\
\ref{fig:projection}, which shows cluster ${\cal A}1$ again at $t=
125\,\tau_{\rm h}$. The function $\Sigma(r)$ is essentially independent
of the projection, and  slight differences occur only where there
are smaller numbers of stars. Analogue to the variations between
different model realisations discussed above, this is the case at very
large and very small $r$, and at separations where wide-binary
depletion occurs.  At large separations, the tidal field breaks the
symmetry. For small $r$ and at the binary gap, it depends on the
projection which separation bin stays populated and which may become
empty.  However, besides these details the overall structure of
$\Sigma(r)$ is projection invariant.

\begin{figure*}[t]
\unitlength1cm
\begin{picture}(16,18.0)
\put(-0.0,-6.4){\resizebox{19.6cm}{!}{\includegraphics{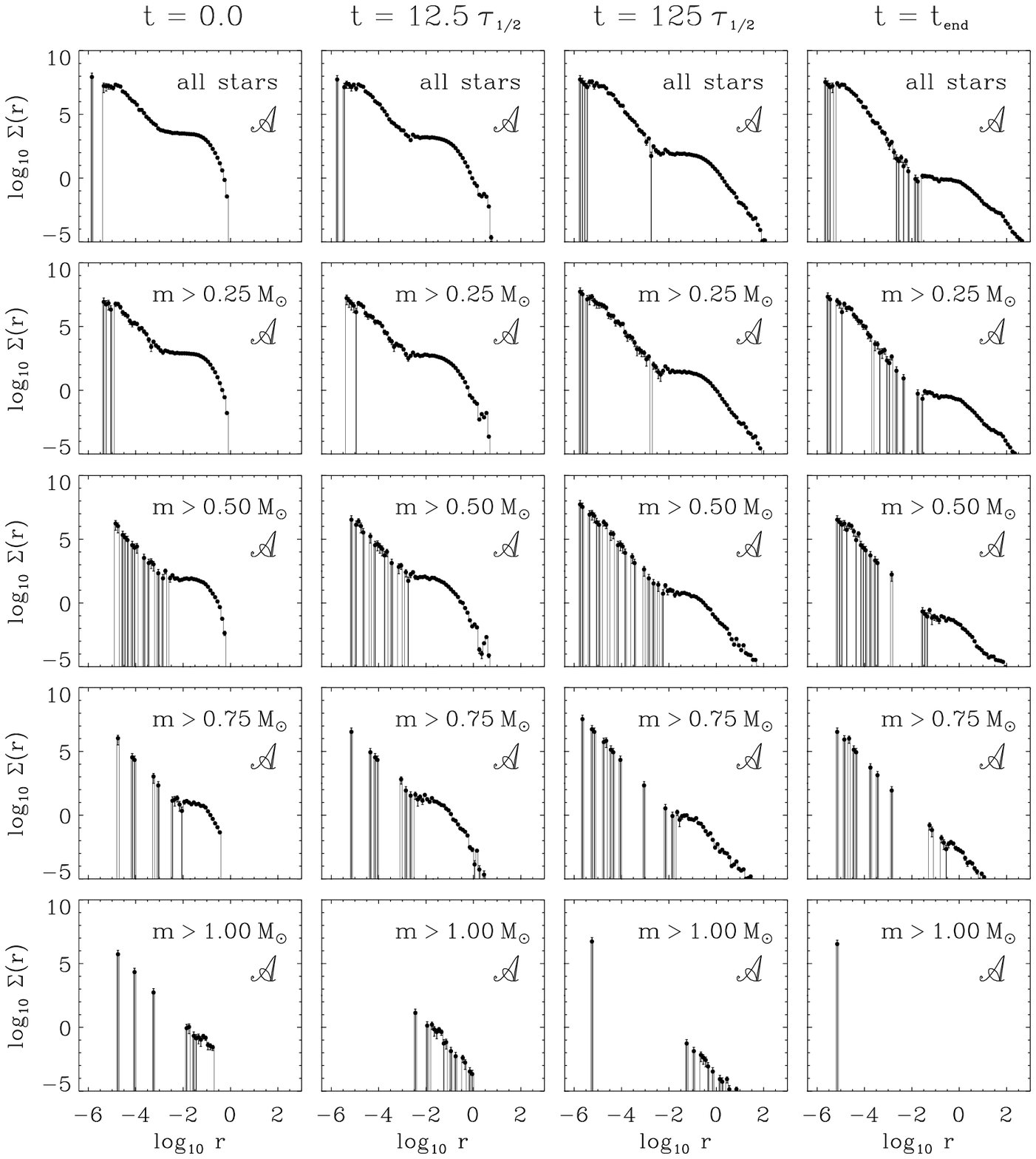}}}
\end{picture}
\hfill
\parbox[t]{\textwidth}{
\caption[fig06]{{\label{fig:mass-cut} Mean surface density of
companions, $\Sigma(r)$, of cluster ${\cal A}$ for different
observational minimum mass limits $m_{\rm min}$. First row of panels:
all stars in the cluster are considered (analogue to top row in figure
\ref{fig:AtoD}). Second row: only stars with $m>0.25\,$M$_\odot$
contribute to $\Sigma(r)$. Third row: $m_{\rm
min}=0.50\,$M$_\odot$. Fourth row: $m_{\rm min}=0.75\,$M$_\odot$, and
lowest row: $m_{\rm min}=1.00\,$M$_\odot$. Times and averaging
procedure are equivalent to figures \ref{fig:AtoD} to
\ref{fig:gap}. There are 400 stars in each cluster realization in the
mass range $0.1\,$M$_\odot$ to $1.1\,$M$_\odot$ following the Kroupa
et al.\ (1993) IMF. The fraction of cluster stars considered in each
panel is $100\,$\%, $51\,$\%, $17\,$\%, $6\,$\%, and $1\,$\%,
respectively.  }}}
\end{figure*}

\section{Observational Bias}
\label{sec:observational-bias}
The most likely observational bias in surveys of stellar clusters
results from the unavoidable detection flux limit, which corresponds
to a stellar mass limit $m_{\rm min}$ such that stars with $m<m_{\rm
min}$ are not detected simply because they are too faint.  This effect
influences the derived mean surface density of companions, as
illustrated in figure\ \ref{fig:mass-cut}, where we again concentrate
on cluster model $\cal A$. As in the previous figures, we plot
$\Sigma(r)$ at four different stages of the cluster evolution,
however, we now `observe' the cluster at various distances, i.e. we
introduce different detection limits $m_{\rm min}$ for the mass. We
consider $m_{\rm min}$ in the range $0.25\,$M$_\odot$ to
$1\,$M$_\odot$.

With increasing $m_{\rm min}$, the total number of detected stars
decreases, and as a result $\Sigma(r)$ is reduced. Also the shape of
$\Sigma(r)$ changes. This effect is small for low detection thresholds
($m\simless0.5\,$M$_\odot$), as the overall star distribution in the
cluster is still well sampled. However, it becomes significant at
large cluster distances when only the brightest stars can be
detected. The binary branch becomes severely under-sampled, and wide
gaps open up. For $m_{\rm min}=1\,$M$_\odot$ the binary branch
disappears in all models.

The inferred cluster core radius also depends quite sensitively on the
completeness of the stellar sample.  As the clusters evolve
dynamically, high-mass stars sink towards the cluster centre due to
mass segregation, whereas low-mass stars move outwards, building up the
extended halo (figure~2 in K3). Sub-populations of higher-mass stars
therefore exhibit smaller core radii as time progresses relative to
the low-mass population.

This trend is clearly seen in figure\ \ref{fig:mass-cut}, where the
second break moves to smaller separations as $m_{\rm min}$ increases.
At very late stages of the dynamical evolution and for very large
cut-off masses, the core radius may become too small, so that the
second break is no longer noticeable.  For example, the function
$\Sigma(r)$ follows an almost perfect $r^{-2}$-power-law for $m_{\rm
min} = 0.75\,$M$_\odot$ at $t_{\rm end}$, exhibiting a smooth
transition from the binary regime to the halo regime without any sign
of the cluster core, which is present when taking all stars into
account. When considering only stars with $m>1\,$M$_{\odot}$, then the
signature of the cluster core disappears at all times. This bias needs
be taken into account when interpreting observational data on star
clusters.

\section{Cluster Morphology}
\label{sec:morphology}

As has been elucidated above, the mean surface density of companions
shows distinct branches, the extend of which appear to couple with
the dynamical state of the cluster. In this section we consider this
in more detail. 

Simple bulk cluster properties that can be used to describe the
dynamical state of a cluster are the core radius, $r_{\rm c}$, the
half-mass radius, $r_{\rm h}$, and the tidal radius, $r_{\rm t}$. The
core radius is approximated by calculating the density-weighted radius
$r_{\rm c}$ (Heggie \& Aarseth 1992),
\begin{equation}
r_{\rm c}^2 = {\sum_{i=1}^{N_{20}} r_{\rm i}^2\,\rho_{\rm i}^2 \over
               \sum_{\rm i=1}^{N_{20}} \rho_{\rm i}^2},
\end{equation}
where $\rho_{\rm i}= 3\,m_{\rm i,5}/(4\pi\,d_{\rm i,5}^3)$ is the
density around star~i estimated within the closest distance $d_{\rm
i,5}$ from star~i containing $n=5$ additional stars with combined mass
$m_{\rm i,5}$, and $r_{\rm i}$ is the distance of star~i from the
density centre of the cluster. The summation extends only over the
innermost 20~\% of all stars in the cluster, $N_{20}$, since this is
sufficient to ensure convergence. The approximate tidal radius (Binney
\& Tremaine 1987),
\begin{equation}
r_t(t) = \left({M_{\rm st}(t)\over3\,M_{\rm gal}}\right)^{1\over3}\,
          r_{\rm GC},
\label{eq:rtid}
\end{equation}
with $M_{\rm gal}=5\times10^{10}\,M_\odot$ being approximately the
Galactic mass enclosed within the distance of the Sun to the Galactic
centre, $r_{\rm GC}=8.5$~kpc. To estimate $r_{\rm t}(t)$, $M_{\rm
st}(t)$ is calculated by summing only those stars which have $r(t)\le
2\,r_{\rm t}(t-\delta t_{\rm op})$, where the data output time
interval $\delta t_{\rm op} \ll t_{\rm relax}(t)$. The quantities
$r_{\rm c}$, $r_{\rm h}$ and $r_{\rm t}$ are averages of $n$ models
per time-snap (Table~\ref{tab:models}).

\begin{figure*}[ht]
\unitlength1cm
\begin{picture}(16, 8)
\put( 0.4, 0.0){\resizebox{16.4cm}{!}{\includegraphics{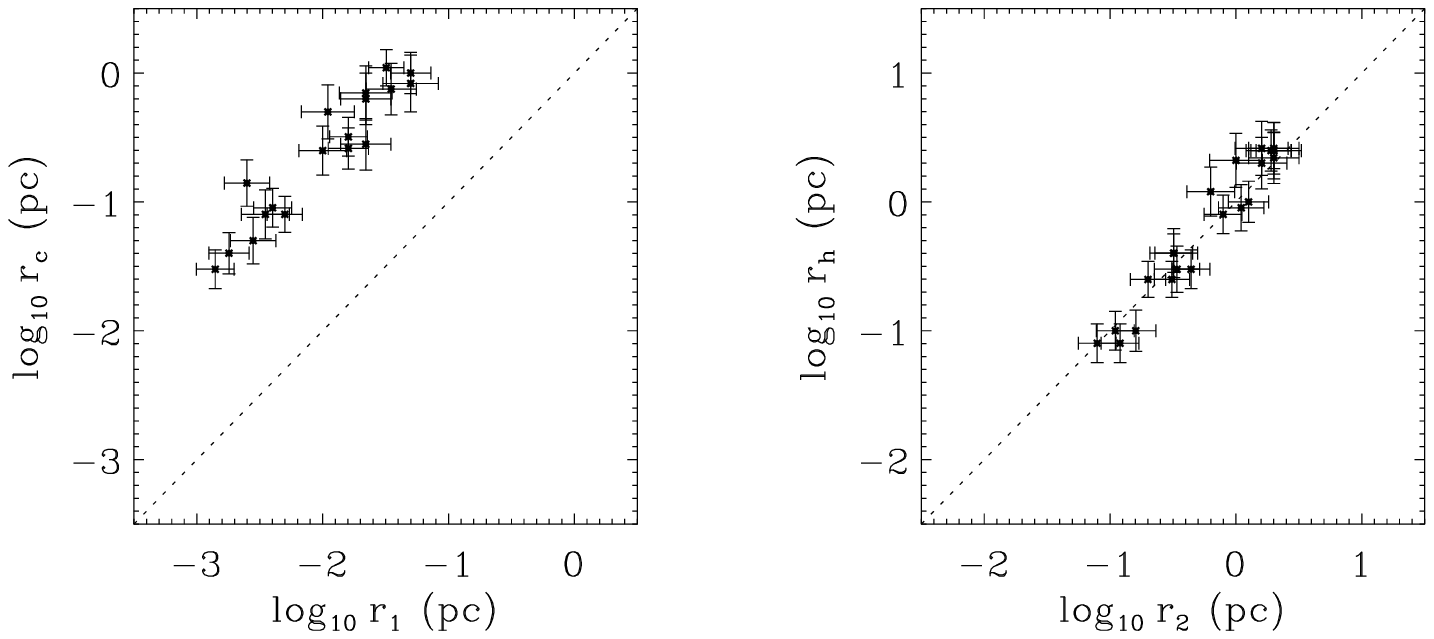}}}
\end{picture}
\hfill
\parbox[t]{\textwidth}{
\caption[fig05] {\footnotesize{\label{fig:corr} Correlation between
the first break radius $r_1$ and cluster core radius $r_{\rm c}$
(left panel), and the radius $r_2$ of the second break and the half
mass radius $r_{\rm h}$ (right panel).}}}
\end{figure*}

We define two breaks in $\Sigma(r)$, $r_1$ and $r_2$, by fitting power
laws to the three distinct branches of $\Sigma(r)$ (the binary branch,
the flat central plateau, and the cluster halo out to the tidal radius
$r_{\rm t}$) and determining the separation at the intersection of the
fits. These values are listed in Table~\ref{tab:radii}, and plots of
$r_{\rm c}$ vs $r_1$ and $r_{\rm h}$ vs $r_2$ are presented in
figure~\ref{fig:corr}.

\begin{table*}[t]
  \begin{center}
    \begin{tabular}[h]{rlcccccc}
\hline
Model & time & $t$ $(10^6\,$yr) & $r_{1}$ (pc) & $r_{2}$ (pc) & $r_{\rm c}$ (pc) & $r_{\rm h}$ (pc) & $r_{\rm t}$ (pc)\\
%&& (pc) & (pc) & (pc) & (pc) & (pc) \\
\hline
Model ${\cal A}$ & $t=t_{\rm init}$      & 0.0  & 0.0014& 0.079& 0.03 & 0.08 & 8.0 \\
                 & $t=12.5 \tau_{\rm h}$ & 1.19 & 0.0018& 0.16 & 0.04 & 0.1  & 8.0 \\
                 & $t=125  \tau_{\rm h}$ & 11.9 & 0.0035& 0.32 & 0.08 & 0.4  & 8.0 \\
                 & $t=t_{\rm end}$       & 300  & 0.011 & 1.0  & 0.5  & 2.1  & 6.0 \\
\hline								              
Model ${\cal B}$ & $t=t_{\rm init}$      & 0.0  & 0.005 & 0.20 & 0.08 & 0.25 & 8.0 \\
                 & $t=12.5 \tau_{\rm h}$ & 6.67 & 0.004 & 0.32 & 0.09 & 0.4  & 8.0 \\
                 & $t=125  \tau_{\rm h}$ & 66.7 & 0.01  & 0.63 & 0.25 & 1.2  & 7.5 \\
                 & $t=t_{\rm end}$       & 300  & 0.022 & 1.6  & 0.7  & 2.6  & 6.2 \\
\hline								              
Model ${\cal C}$ & $t=t_{\rm init}$      & 0.0  & 0.016 & 0.79 & 0.32 & 0.8  & 8.0 \\
                 & $t=12.5 \tau_{\rm h}$ & 37.5 & 0.016 & 1.26 & 0.26 & 1.0  & 8.0 \\
                 & $t=t_{\rm end}$       & 300  & 0.035 & 2.0  & 0.75 & 2.6  & 6.7 \\
\hline								              
Model ${\cal D}$ & $t=t_{\rm init}$      & 0.0  & 0.032 & 2.0  & 1.1  & 2.5  & 8.0 \\
                 & $t=12.5 \tau_{\rm h}$ & 211  & 0.05  & 1.9  & 1.0  & 2.5  & 7.2 \\
                 & $t=t_{\rm end}$       & 296  & 0.05  & 2.0  & 0.83 & 2.5  & 6.7 \\
\hline								              
Model ${\cal E}$ & $t=t_{\rm init}$      & 0.0  & ---   & 0.12 & 0.03 & 0.08 & 8.0 \\
                 & $t=12.5 \tau_{\rm h}$ & 1.19 & ---   & 0.11 & 0.03 & 0.1  & 8.0 \\
                 & $t=125  \tau_{\rm h}$ & 11.9 & 0.0028& 0.34 & 0.05 & 0.3  & 8.0 \\
                 & $t=t_{\rm end}$       & 297  & 0.022 & 1.6  & 0.28 & 2.0  & 6.4 \\
\hline								              
Model ${\cal F}$ & $t=t_{\rm init}$      & 0.0  & ---   & 0.31 & 0.07 & 0.25 & 8.0 \\
                 & $t=12.5 \tau_{\rm h}$ & 6.67 & ---   & 0.44 & 0.07 & 0.3  & 8.0 \\
                 & $t=125  \tau_{\rm h}$ & 66.7 & 0.0025& 1.1  & 0.14 & 0.9  & 7.8 \\
                 & $t=t_{\rm end}$       & 300  & 0.022 & 2.0  & 0.63 & 2.2  & 6.9 \\
\hline
    \end{tabular}
    \end{center}
    \hfill
    \vspace{0.2cm}
    \parbox[b]{\textwidth}{
    \caption[tab2]{
    \label{tab:radii} 
Morphologically important length scales. The table lists the separations
$r_1$ and $r_2$, where the first and second break of the
distribution $\Sigma(r)$ occur in all model clusters. The values for
$r_{1}$ and $r_2$ are obtained at the intersection of the power-law
fits to the binary branch and the central flat plateau, and the
plateau and the halo distribution of cluster stars out to the tidal
radius, respectively. These separations are compared to the core
radius $r_{\rm c}$ and the half-mass radius $r_{\rm h}$ at the
different times in Fig.~\ref{fig:corr}.}}
\end{table*}

The figure shows that the quantities are well
correlated. Specifically, we find (the uncertainties are mean absolute
deviations)
\begin{equation}
\langle r_{\rm c}/r_1 \rangle = 20.5 \pm 9.7
\label{eqn:corr1}
\end{equation}
and 
\begin{equation}
\langle r_{\rm h}/r_2 \rangle = 1.14 \pm 0.29.
\label{eqn:corr2}
\end{equation}
This suggests that the break points in $\Sigma(r)$ can be used to
infer the core radius and half-mass radius of a cluster.  We also find
good correlations between $r_{\rm h}$ and $r_{\rm c}$ ($\langle r_{\rm
h}/r_{\rm c} \rangle = 3.86 \pm 1.04$) and between the break points in
$\Sigma(r)$ ($r_2/r_1 = 74.3 \pm 44.4$).

The above scaling relations apply for the specific low-mass cluster
models that we investigate. These models do not suffer significant
core collapse, which is partly given by the relatively fast
evaporation time ($\approx0.5-1$~Gyr, K3), and the ubiquitous binary
stars which oppose core collapse.  More massive clusters are likely to
show different correlations, notably between $r_{\rm h}$ and $r_{\rm
c}$, and between $r_1$ and $r_2$, since core collapse leads to the
contraction of $r_{\rm c}$ but an expansion of $r_{\rm h}$ (Giersz \&
Spurzem 2000).  Analysis of more massive clusters using $\Sigma(r)$ is
a future goal, and it will be interesting to see if
correlations~\ref{eqn:corr1} and~\ref{eqn:corr2} remain valid.

\section{Summary}
\label{sec:summary}
In this paper, we investigate how dynamical cluster evolution is
manifest in the mean surface density of companions, $\Sigma(r)$, as a
function of separation $r$. We find that throughout all evolutionary
phases, $\Sigma(r)$ can be subdivided into four distinct branches,
each following approximately a power-law behaviour:

\begin{enumerate}

\item At small separations, and throughout all evolutionary phases of
the star cluster, $\Sigma(r)$ traces binary stars and higher-order
multiple systems. In the binary branch the slope of the companion
density is approximately $-2$.

The binary population keeps a memory of the dynamical evolution of
star clusters.  The depletion of the binary population at large
separations leads to a steepening of $\Sigma(r)$ in the wide-binary
branch. In extreme cases, it may become completely depopulated and a
gap between binaries and the cluster branch opens up.  The gap
increases with cluster age and with increasing initial density. For
the binary branch, $\Sigma(r)$ is sensitive to the initial properties
of the system, so that analysing $\Sigma(r)$ for a cluster can
constrain its birth configuration.

\item The well-defined binary branch blends into a plateau of constant
companion density which corresponds to the main body of the stellar
cluster, and which we refer to as the cluster branch in $\Sigma(r)$.
The transition occurs at separations when chance projections of
cluster members begin to outnumber the contribution of binary stars in
that separation bin. The location of this {\em first} break at $r_1$
depends on the binary fraction and the central stellar density of the
cluster, and thus on $r_{\rm c}$.  As the cluster expands, the density
decreases and the core radius increases with $r_{\rm c}\approx
21\,r_1$ for the set of models studied here.  As a consequence, the
binary branch becomes less affected by crowding and the first break in
$\Sigma(r)$ shifts to greater separations.  Generally, as the cluster
evolves the plateau `decreases in height' and `moves' to larger
separations. Larger parts of the binary branch are `uncovered' as a
result of the expansion and the formation of wide hierarchical systems
through capture.
  
\item Relaxed Galactic clusters exhibit a core/halo structure. This
becomes apparent in $\Sigma(r)$ through the existence of a second
break at $r_2$, which is approximately located at the half-mass
radius, $r_{\rm h}\approx r_2$, beyond which $\Sigma(r)$ decreases
again. The slope of this halo branch depends sensitively on the
evolutionary state of the cluster.  If the cluster is still
sufficiently young so that most of it is confined well inside it's
tidal radius, then $\Sigma(r)$ decreases rapidly. However, relaxed
clusters that fill their tidal radii have a halo branch with a slope
of approximately $-2$.
  
\item Beyond the tidal radius, the density of unbound stars decreases
gradually, which is reflected in a fourth branch in $\Sigma(r)$. It is
a measure of the density distribution of the moving group relative to
the centre of its origin, which is the star cluster. The decay of
$\Sigma(r)$ is steeper here because the stars are finally removed from
the vicinity of the cluster within an orbital period about the Galaxy
(Terlevich 1987).

\item The mean surface density of companions closely reflects
morphological properties of stellar clusters. The break points in
$\Sigma(r)$, $r_1$ and $r_2$,  therefore can be used to infer the core
radius $r_{\rm c}$ and half-mass radius $r_{\rm h}$ of a
cluster. These quantities are well correlated, and we find $\langle
r_{\rm c}/r_1 \rangle = 20.5 \pm 9.7$ and $\langle r_{\rm h}/r_2
\rangle = 1.14 \pm 0.29$, respectively.

\end{enumerate}

Altogether, $\Sigma(r)$ contains valuable information on the dynamical
state of a star cluster, and allows an assessment of cluster
properties such as the core and half-mass radii.  As the location of the
first break in $\Sigma(r)$ (or the possible occurrence of a gap between
binary and cluster branch) depends on the binary fraction and on the
initial central density, these parameters can in principle be
inferred from analysing $\Sigma(r)$. 

Our study confirms that different projections of the same data do not
change $\Sigma(r)$ significantly during the evolution of initially
spherical clusters in the Galactic tidal field. Also, different
numerical renditions of the same models lead to indistinguishable
results. Hence, they can be combined to improve the statistical
significance of the ensemble average $\Sigma(r)$.

To allow for a proper assessment of stellar cluster properties using
$\Sigma(r)$, it is important to consider as complete a census of
cluster stars as possible. Observational bias (i.e. non-detection of
faint stars) may complicate the interpretation of $\Sigma(r)$ and
limit its applicability for inferring the initial stage of the cluster
under study (see also Bate et al.\ 1998).  Our tests show, however,
that $\Sigma(r)$ remains a useful quantity even when stars with
$m\simless0.5\,$M$_\odot$ are not detected. We also find that mass
segregation is evident in $\Sigma(r)$ through the location of the
second break in dependence of the mass-range used to construct
$\Sigma(r)$.

Future analysis of numerical models of rich clusters, for which mass
segregation and possibly core collapse play important roles in the
late phases of the dynamical evolution, will be performed to deepen
the issues raised in this pilot study.

\begin{acknowledgements}
We thank Sverre Aarseth for distributing {\sc Nbody5} freely. RSK
acknowledges support by a Otto-Hahn-Stipendium from the
Max-Planck-Gesellschaft and partial support through a NASA
astrophysics theory program at the joint Center for Star Formation
Studies at NASA-Ames Research Center, UC Berkeley, and UC Santa Cruz.
PK acknowledges support from DFG grant KR1635.
\end{acknowledgements}

\end{document}